\documentclass[twocolumn,a4paper,11pt]{article}
\usepackage{blindtext}
\usepackage{titlesec}
\usepackage[utf8]{inputenc}
\usepackage{times}
\usepackage{microtype}

\usepackage{lscape}

\usepackage{epstopdf}

\usepackage[numbers,sort&compress]{natbib}
\usepackage{subfigure}
\usepackage{multirow}
\usepackage{float}
\usepackage{soul}
\usepackage{xcolor}
\graphicspath{{./images/}}

\usepackage{amssymb}
\usepackage{amsmath}
\usepackage{mathtools}
\usepackage{amsthm}
\usepackage{bm}

\usepackage{appendix}

\providecommand{\keywords}[1]
{
  \small	
  \textbf{\textit{Keywords---}} #1
}

\bibpunct[, ]{[}{]}{,}{}{,}{,}
\renewcommand\bibfont{\fontsize{10}{12}\selectfont}

\usepackage{authblk}
\title{Correlated functional models with derivative information for modeling MFS data on rock art paintings}
\author[1,$\ast$]{Gabriel Riutort-Mayol}
\author[2]{Virgilio G\'omez-Rubio}
\author[1]{Jos\'e Luis Lerma}
\author[3]{Julio M. del Hoyo-Mel\'endez}
\affil[1]{Department of Cartographic Engineering, Geodesy, and Photogrammetry, Universitat Polit\`ecnica de Val\`encia, Spain}
\affil[2]{Department of Mathematics, Universidad de Castilla-La Mancha, Spain}
\affil[3]{Laboratory of Analysis and Non-Destructive Investigation of Heritage Objects, The National Museum in Krakow, Poland}
\affil[$\ast$]{gabriuma@gmail.com}
\date{}                     
\begin{document}

\maketitle

\begin{abstract}
Microfading Spectrometry (MFS) is a method for assessing light sensitivity color (spectral) variations of cultural heritage objects. Each measured point on the surface gives rise to a time-series of stochastic observations that represents color fading over time. Color degradation is expected to be non-decreasing as a function of time and stabilize eventually. These properties can be expressed in terms of the derivatives of the functions. In this work, we propose spatially correlated splines-based time-varying functions and their derivatives for modeling and predicting MFS data collected on the surface of rock art paintings. The correlation among the splines models is modeled using Gaussian process priors over the spline coefficients across time-series. A multivariate covariance function in a Gaussian process allows the use of trichromatic image color variables jointly with spatial locations as inputs to evaluate the correlation among time-series, and demonstrated the colorimetric variables as useful for predicting new color fading time-series. Furthermore, modeling the derivative of the model and its sign demonstrated to be beneficial in terms of both predictive performance and application-specific interpretability.
\end{abstract}


\section{Introduction}

\subsubsection*{Case study and motivation}

Prehistoric rock art paintings (see, for example, Figure \ref{observed_curves}) are exposed to environmental elements, which can accelerate their degradation, increasing the risk of losing such a valuable cultural piece from past societies. A part from and in addition to many other factors, exposure to sunlight can have adverse effects on these systems due to thermal and photochemical degradation of the historic materials \citep{Diez-Herrero_2009}. 

This is a surface effect and the degree of color change depends on the chemical composition of the pigment. In this study we focused on the study and documentation of the degree of color changing/fading of paintings, patinas and host rock which is crucial for the conservation of these systems \citep{Cassar_2001, del_hoyo_2015}.

Materials with higher light sensitivity usually experience a rapid color change during the early stages of exposure, followed by a slower rate after maximum fading has occurred, assuming total disappearance of the substance that produces the color (chromophore) at this second stage of the fading \citep{feller1986determination,giles_1965,giles_1968,johnston1984kinetics}. Thus, color fading is expected to be non-decreasing as a function of time and saturate in the long term. 

Microfading Spectrometry (MFS) is a method for assessing light sensitivity color (spectral) variations of cultural heritage objects. This method allows for real time monitoring of spectral changes of a cultural heritage object by undertaking in situ lightfastness measurements on the surface under study \citep{Whitmore_1999,Whitmore_2000,del_hoyo_2010}. MFS technique provides for each measured point on the surface of rock art paintings a time-series of observations that represent potential color fading along time to exposition to light \citep{Whitmore_1999, del_hoyo_2010}. Thus, MFS measurements can be seen as observations of an underlying spatio-temporal stochastic process.

The MFS instrument is very sensitive to movement and glossy surface effects, causing extremely large fluctuations in the measurements. Furthermore, they can be easily contaminated by changes in the lighting conditions when the measurements are performed in outside environments. These large fluctuations and possible systematic noise effects in the observations can cause that models do not satisfy those properties of monotonicity and long-term stabilization of color fading over time.
  
Existing lightfastness studies on these systems have been limited to analyze a few measured points on the surface of the rock art paintings because these measurements are largely time-consuming and difficult to materialize under harsh conditions. So, in the paper we propose an interpolation procedure in order to extend the analysis to others unobserved locations on the surface of rock art paintings. A complete map of estimated color fading data for the entire surface under study is an important and useful information in order to achieve further successful conservation actions.

\subsubsection*{Background methods}
Functional data usually refers to independent realizations of a functional random variable that takes values in a continuous space. Time-series of observations (e.g. color-fading time-series functions) might be the most common case of functional data in 1D, $f(x):x\in {\rm I\!R} \to {\rm I\!R}$, but spatially distributed observations can also be seen as functional data in a 2D space, $f(x):x\in {\rm I\!R}^2 \to {\rm I\!R}$, or spatio-temporal observations can also be seen as functional data in a 3D space, $f(x):x\in {\rm I\!R}^3 \to {\rm I\!R}$.

In order to construct a model useful for making predictions of new functional data as function of new values of the variables in the input space, the process must be considered as an structured process with dependence among observations. 

Correlated functional models consider the observed functional data as non-independent functions. A popular approach for correlated space-temporal functional data consists in three-way (spatial (2D) and temporal (1D)) penalized splines models \citep{wood2003thin} with different basis constructions based on Kronecker products \citep{currie2006generalized,lee2011p} or additive basis components \citep{kneib2006structured}. \cite{aguilera2017prediction}  propose a mixture of functional regression model for functional response and penalized spline spatial regression. However, in general, these models become highly parameterized, hardly interpretable, and difficult to fit specially in a Bayesian framework and using sampling methods. 

Another powerful approach consists in considering the space-time structured observations as stochastic realizations of a Gaussian process prior with a spatio-temporal covariance function. Gaussian processes (GPs) \citep{Neal_1999, Rasmussen_2006} are a natural and flexible non-parametric prior models for $D$-dimensional (e.g. space and time) functions with multivariate predictors (input variables) in each dimension. In a separable form, the space-time covariance function is a result of the interaction of the two independent processes, space and time \citep{banerjee_2014}. In a non-separable form, the covariance function models space-time interaction \citep{cressie1999classes,de2002nonseparable}. The drawback of GP models is their expensive computational demand in covariance matrix inversion that is in general a $O\big((NT)^3\big)$ operation, with $N$ the number of spatial $(S_x,S_y)$ locations and $T$ the number of time points.

A geostatistical approach for spatio-temporal data is the kriging approach for 1D-functional data \citep{Giraldo_2010}, which the 1D-functional data consists in the time-series of the data. The spatial correlation of the time-series is modeled in the covariance function. The dimension of the covariance matrix is $N\times N$ and the matrix inversion is a $O(N^3)$ operation, with $N$ the number of spatial locations. This approach although requires less computation that the spatio-temporal GPs, has the drawback of being a quite less flexible model in the spatio-temporal structure since the same spatial structure is defined for the whole time-series. A related approach can be found in \cite{Baladandayuthapani_2008} where the spatial correlation between the time-series is modeled by defining GPs with a spatial covariance function across the time-series functional coefficients. This construction allows for modeling different spatial structure for the different orders of the coefficients.

Furthermore, inducing monotonicity or gradient to functions can be expressed in terms of the derivative of the function, as either the sign of a derivative for monotonicity or the value of a derivative for gradients. As diferentation is a linear operator, the derivative of both semiparametric models and Gaussian process models can be used as additional observations jointly with the regular observations in order to force the function to fit these properties \citep{rasmussen_2003, shively_2009}. \cite{Riihimaki_2011} uses a Gaussian process model and the information of its derivative process to induce monotonicity on the functions using virtual observations of the sign of the derivatives. 

On the other hand, monotonicity in semiparametric models can be forced by construction. \cite{shively_2009} proposes two approches to obtain monotonic functions, the first using a modified characterization of the smooth monotone functions proposed in \cite{ramsay_1998} that allows for unconstrained estimation, and the second using constrained prior distributions for the regression coefficients to ensure the monotonicity.  \cite{Brezger_2008} induces monotonicity on semiparametric and non-parametric models imposing the restriction that the regression coefficients are ordered, also sometimes known as isotonic regression. These constraints are imposed by specifying truncated prior distributions in order to reject the undesired draws for the parameters in the MCMC sampling. \cite{Reich_2011} makes a similar approach imposing order to the regression parameters by means of reparameterizing and constraining these parameters with application to a quantile regression model.

\subsubsection*{Proposed methodology}
In this paper, a specific application with the aim of modeling and predicting color fading time-series for new unobserved spatial locations on the surface of rock art paintings is carried out. 

We consider spatially correlated functional models, where the functional models correspond to the color fading time-series.  The derivative process of the functional models to ensure monotonicity (non-decreasing) and long term stabilization of color fading (long term color fading \citep{del_hoyo_2015}) as a function of time, is taken into account. 

The color-fading time-series are modeled as penalized splines-based models. The derivative of a semiparametric model such as a splines model is still a linear model and can be used as a additional observation jointly with the regular observations. Virtual derivative observations of both the values of the partial derivatives and the sign of the partial derivatives are used to induce monotonicity and long-term saturation, respectively.

The spatial correlation of the time-series is modeled by defining multivariate Gaussian process priors distributions over the splines coefficients across time-series. Color fading is mainly related to physicochemical properties of the measured surface. Physicochemical data for all the points on the surface to predict new time-series are hard to obtain. Instead, image color values can be used as input variables to construct and evaluate the correlation among time-series, since these image color variables are related in some way to physicochemical properties of the imaged surface \citep{Malacara_2011}. A multivariate covariance function in a Gaussian process prior allows for using trichromatic image color variables jointly with spatial locations as inputs to evaluate the covariance structure among time-series.



Finally, in order to do model evaluation and comparison, the same model but without derivative information is also fitted. Cross-validation procedures are conducted to compute the \textit{posterior predictive checks}, the \textit{expected log predictive density} and the \textit{mean square predictive errors} in order to do model checking and selection and evaluate the predictive performance.


The rest of the paper is structured as follows. Section \ref{sec:data} describes the available data in detail. Section \ref{sec:model} focuses on the modeling and inference formulation, as well as model checking and model selection procedures. Section \ref{sec:results} describes the results of applying the proposed model to the data set. Section \ref{discussion} discusses about the results. Finally, Section \ref{conclusion} presents a brief conclusion of the work.

\section{Data description} \label{sec:data}
The MFS technique allows for real time monitoring of spectral changes of a small area (diameter of 0.5 mm) of a cultural heritage object by exposing it to light. This method was developed by \cite{Whitmore_1999,Whitmore_2000} and has been extensively used to study the lightfastness of cultural objects \citep{Ford_2011,Ford_2013,Columbia_2013}. 

\begin{figure*}
\centering
\subfigure{\includegraphics[width=3in, trim = 10mm 10mm 10mm 0mm, clip]{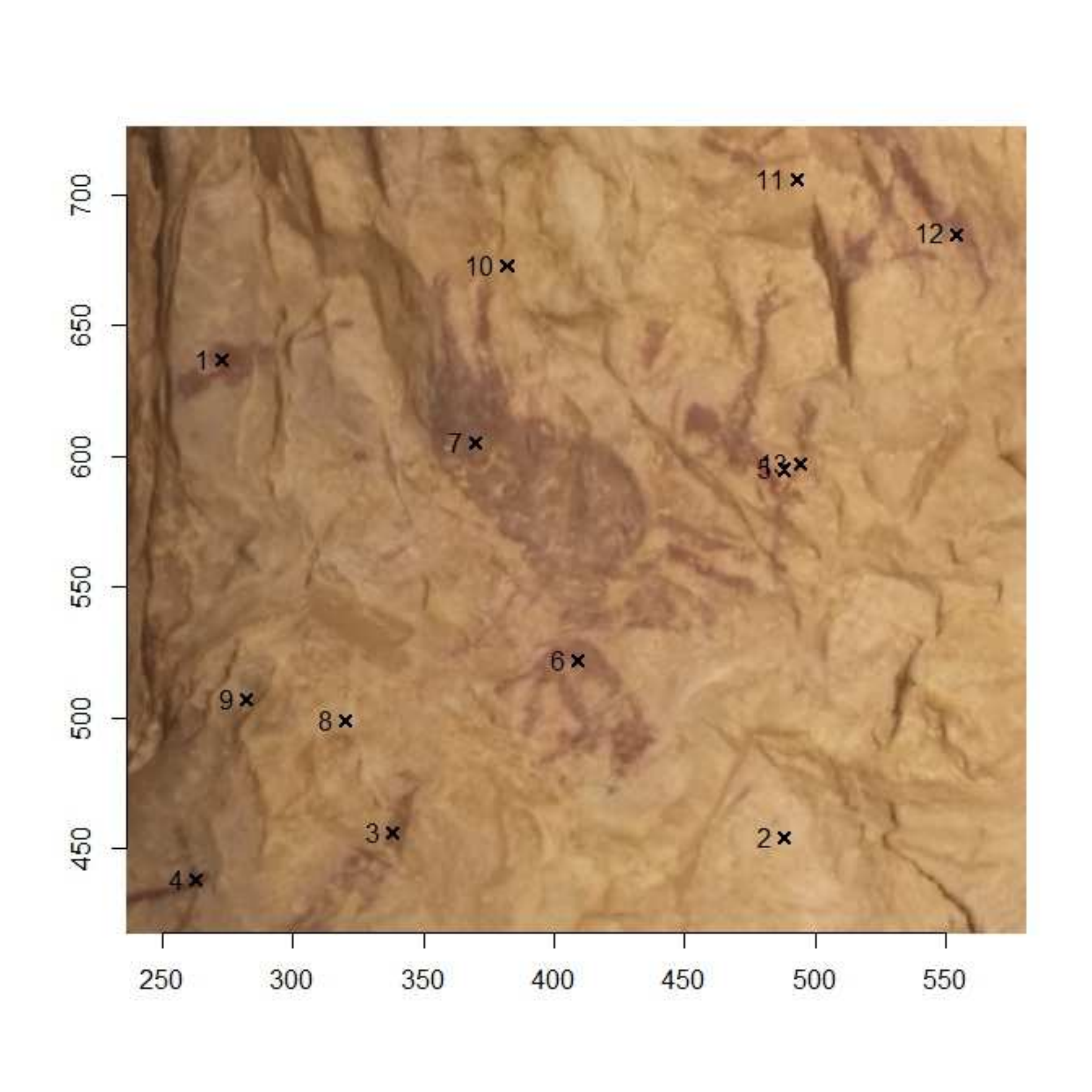}}
\hspace{0.2cm}
\subfigure{\includegraphics[width=3in, trim = 0mm 1mm 0mm 0mm, clip]{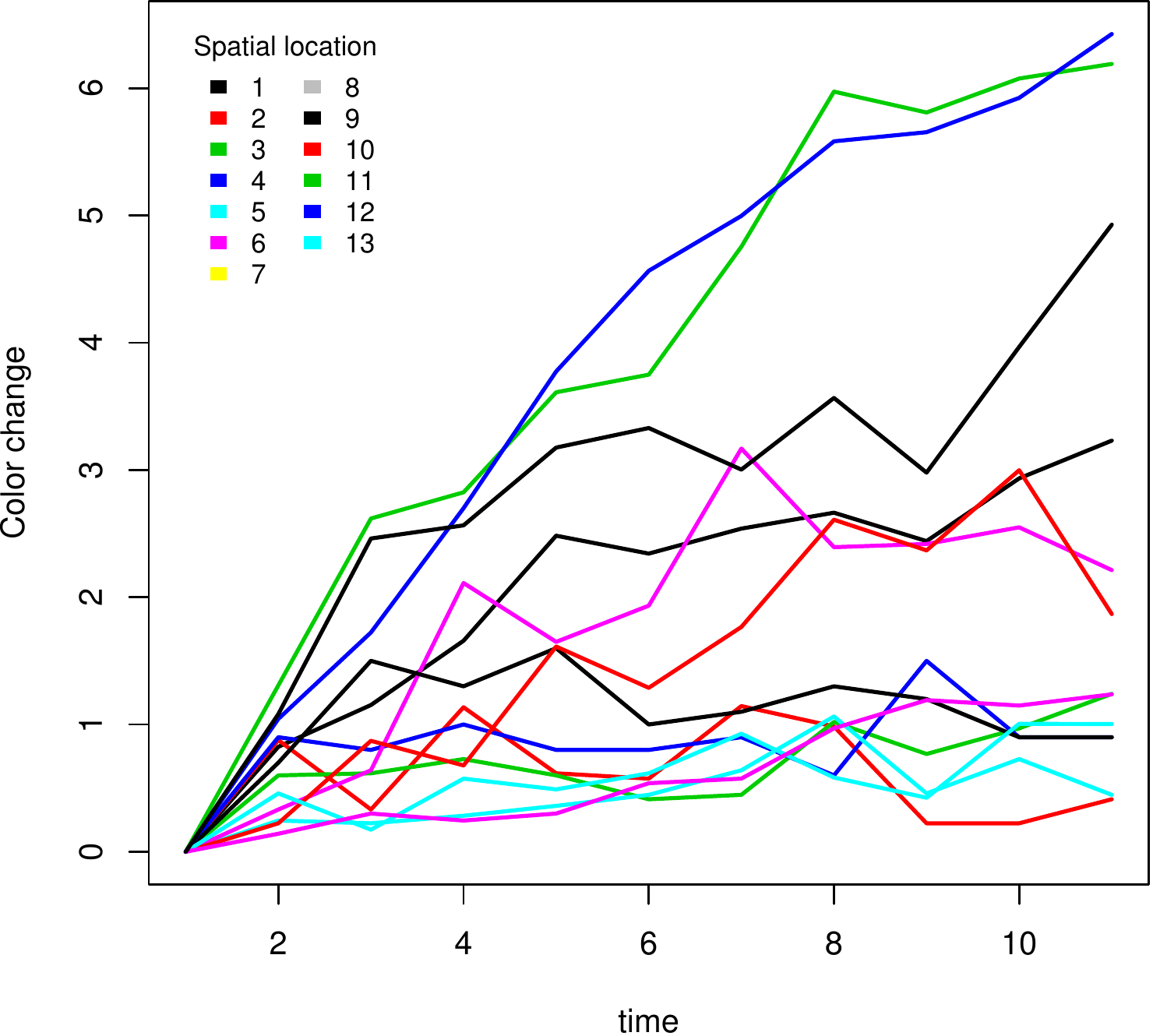}}
\caption{(left) Image with the points where the MFS observations were measured in Shelter V of Cova Remigia. (right) Observed MFS time-series.}
  \label{observed_curves}
\end{figure*}

In this practical case we seek to evaluate and document the degree of color fading over time and space on rock art paintings caused by direct solar irradiation. Each measured point on the surface gives rise to a time-series that represents color fading (Color differences/Delta E* value, \cite{Malacara_2011}) over time. The MFS measurements have a duration of 10 minutes. The sampling frequency will be once per minute, such that the resulting time-series will consist of 11 ($T=11$) time points $(t=1,\dots,11)$. 
Due to these measurements are largely time-consuming and difficult to materialize, specially in these rock art systems, the MFS dataset available consisted of 13 observed locations on the surface ($N=13$) only. Figure \ref{observed_curves}-left shows their pixel locations on a color image of the study area; each location incorporates color fading time-series of observations (Figure \ref{observed_curves}-right). The rock art paintings study area is located in the Cova Rem\'igia rock-shelter, Castell\'on (Spain).

The number of available input variables to evaluate the covariance between spatial locations is 5 ($D=5$), which are the three image color variables, Hue ($H$), Saturation ($S$) and Intensity ($I$), and the two spatial coordinates $S_x$ and $S_y$. Table \ref{tab:summary} contains summary statistics for these input variables, which have been rescaled by dividing by their standard deviations. The $S_x$ and $S_y$ spatial coordinates were divided by their common standard deviation.

Each time-series is modeled as a spline-based function. 
The order or number of spline knots $K$ of the spline-based functions model is set to 3 ($k=1,\dots,K$) and placed uniformly through the time points variable. 

\begin{table}
\caption{\label{tab:summary}Summary statistics of the input variables}
\centering
\fbox{
\begin{scriptsize}
\begin{tabular}{l|ccccc}
  & H & S & I & $S_x$ & $S_y$ \\
\hline
Mean & 5.255 & 9.704 & 5.155 & 3.549 & 4.969 \\
Std. Dev. & 1.0 & 1.0 & 1.0 & 0.732 & 0.674 \\
1st Qu. & 4.359 & 9.042 & 4.603 & 2.910 & 4.387 \\
3rd Qu. & 5.913 & 10.273 & 5.772 & 4.187 & 5.551 \\
\end{tabular}
\end{scriptsize}
}
\end{table}

\section{Correlated time-series with derivative information}\label{sec:model}

\subsubsection*{Penalized splines models}
We consider a continuous stochastic process based on time-series of observations observed at spatial locations on a surface.  
Thus, $y_{ti}$ denotes the $t$th value of the time-series at the $i$th location, with $t = 1,..,T$ representing the time points and $i = 1,..,N$ representing the spatial locations. The time-varying dimension of the data is modeled using basis functions models, so the $T$-vector of values of a time-series $i$, $\bm{y}_{\bm{\cdot} i}=(y_{1i},\dots,y_{Ti})^\top \in {\rm I\!R}^T$, is considered to follow a Normal distribution depending on an underlying penalized quadratic-splines-based function $\bm{f}_{\bm{\cdot} i}:{\rm I\!R}^T \rightarrow (f_{1i},\dots,f_{Ti})^\top \in {\rm I\!R}^T$ and noise variance $\sigma^2$. This can be expressed as:
\begin{equation*}
p(\bm{y}_{\bm{\cdot} i} | \bm{f}_{\bm{\cdot} i}, \sigma^2) = \mathcal{N}(\bm{y}_{\bm{\cdot} i} | \bm{f}_{\bm{\cdot} i},\sigma^2 I)
\end{equation*}
\begin{equation}\label{eq_fsplines}
	\bm{f}_{\bm{\cdot} i} = H \bm{\beta}_{\bm{\cdot} i} + W \bm{b}_{\bm{\cdot} i}
\end{equation}
  
\noindent In the previous equation, $H \bm{\beta}_{\bm{\cdot} i}$ is the linear part of the model for the time-series $i$, where $H \in {\rm I\!R}^{T\times 2}$ is the matrix of linear functional values with the $t$th row $\bm{H}_{t \bm{\cdot}}=(1,t)$, and $\bm{\beta}_{\bm{\cdot} i}=(\beta_{1i},\beta_{2i})^\top$ are the linear coefficients for the time-series $i$. On the other hand, $W \bm{b}_{\bm{\cdot} i}$ is the non-linear or splines-based part of the model for time-series $i$, where $\bm{b}_{\bm{\cdot} i}=(b_{1i},\dots,b_{Ki})^\top \in {\rm I\!R}^K$ are the splines coefficients for time-series $i$, and $K$ the order of the splines model and number of knots, and $W \in {\rm I\!R}^{T\times K}$ is the matrix  of penalized quadratic-spline functions values:
\begin{equation}\label{eq_W}
W= Z \cdot \Omega^{-1/2},
\end{equation}

\noindent where $Z \in {\rm I\!R}^{T\times K}$ is the matrix of the quadratic-spline functions values, and $\Omega^{-1/2}\in {\rm I\!R}^{K\times K}$ the matrix of penalization of \cite{Crainiceanu_2005}. An element $(t,k)$ of the matrix $Z$ is
\begin{equation}\label{eq_Z}
Z_{tk}=(t-\kappa_k)^2,
\end{equation}

\noindent and of the matrix $\Omega$ is $\Omega_{lk}=(\kappa_l - \kappa_k)^{2}$, with $l,k=1,...,K.$ $\kappa_l$ and $\kappa_k$ are the $l$th and $k$th pre-fixed knots corresponding to the $l$th and $k$th spline function, respectively.
\\

The observational model for the matrix of observations $\bm{y}=(\bm{y}_{\bm{\cdot} 1},\dots,\bm{y}_{\bm{\cdot} N}) \in {\rm I\!R}^{T\times N}$ takes the form
\begin{equation*}
p(\bm{y}|\bm{f},\sigma^2) = \prod_{\forall t,i} \mathcal{N}(y_{ti}|f_{ti},\sigma^2),
\end{equation*}

\noindent with
\begin{eqnarray}
&\bm{f}=H \bm{\beta} + W \bm{b} \\
& f_{ti}=\bm{H}_{t \bm{\cdot}} \bm{\beta}_{\bm{\cdot} i} + \bm{W}_{t \bm{\cdot}} \bm{b}_{\bm{\cdot} i} \label{eq_f}
\end{eqnarray}

\noindent In the previous equations $\bm{\beta} \in  {\rm I\!R}^{2\times N}$ is the matrix of linear coefficients with the $i$th column $\bm{\beta}_{\bm{\cdot} i}=(\beta_{1i},\beta_{2i})^\top$, $\bm{b}=(\bm{b}_{\bm{\cdot} 1},\dots,\bm{b}_{\bm{\cdot} N}) \in {\rm I\!R}^{K\times N}$ is the matrix of splines coefficients  with $i$th column $\bm{b}_{\bm{\cdot} i}=(b_{1i},\dots,b_{Ki})^\top \in {\rm I\!R}^K$. $H$ and $W$ are the matrices of linear and splines function values, respectively, and $\bm{W}_{t \bm{\cdot}}=(W_{t1},\dots,W_{tK}) \in {\rm I\!R}^K$ and $\bm{H}_{t \bm{\cdot}}=(1,t) \in {\rm I\!R}^2$ are the $t$th rows of the matrices $W$ and $H$, respectively. 
\vspace{0.3cm}

\subsubsection*{Correlating the splines models}
In order to establish a correlation structure among time-series $i$, the matrix of spline coefficients $\bm{b} \in {\rm I\!R}^{K\times N}$ is considered as a realization of a continuous stochastic process, modeled as a (zero-mean) Gaussian process prior.

With the aim of simplify the covariance structure, null covariances between splines coefficients belonging to different spline knots can be considered, that is equivalent to define $K$ independent Gaussian process priors, one for each $k$th row $\bm{b}_{k \bm{\cdot}}=(b_{k1},\cdots,b_{kN}) \in {\rm I\!R}^{N} $ of matrix $\bm{b}$, i.e.,
\begin{eqnarray*}
p(\bm{b}_{k \bm{\cdot}}|X,\bm{\rho}_k) = \mathcal{N}\big(\bm{b}_{k \bm{\cdot}} | 0,\alpha_k C_k(X;\bm{\rho}_k)\big),
\end{eqnarray*}

\noindent for $k=1,\dots,K$. $C_k$ is the covariance function for the vector of coefficients $\bm{b}_{k \bm{\cdot}}$. Hyperparameter $\alpha_k$ is the standard deviation of the Gaussian process which controls the overall scale or magnitude of the range of values of the vector $\bm{b}_{k \bm{\cdot}}$. Thus, specific covariance structure for each $k$th vector of splines coefficients $\bm{b}_{k \bm{\cdot}}$ can be specified. 

Furthermore, in this work we will simplify even more the covariance structure considering the same spatial structure for every vector of splines coefficients $\bm{b}_{k \bm{\cdot}}$, i.e.,
\begin{eqnarray} \label{eq_gpprior}
p(\bm{b}_{k \bm{\cdot}}|X,\bm{\rho}) = \mathcal{N}\big(\bm{b}_{k \bm{\cdot}} | 0,\alpha_k \, C(X;\bm{\rho})\big),
\end{eqnarray}

\noindent for $k=1,\dots,K$, and $C$ is a common covariance function for every vector of coefficients $\bm{b}_{k \bm{\cdot}}$. 

The $N \times N$ covariance matrix $C$ is computed by a squared exponential covariance function \citep{Rasmussen_2006}, dependent on the matrix $X \in {\rm I\!R}^{N\times D}$ of input variables, where $D$ is the number of inputs variables, and the vector of lengthscale hyperparameters $\bm{\rho}=\{\rho_d\}_{d=1}^{D}$,
\begin{eqnarray*} 
C(X;\bm{\rho})_{(i,j)}=\exp\left( -\frac{1}{2} \, \sum_{d=1}^{D}\frac{1}{\rho^2_d}(X_{id}-X_{jd})^2\right),
\end{eqnarray*}

\noindent where $i,j=1,\cdots,N$. The vector $\bm{\rho}=\{\rho_d\}_{d=1}^{D}$ contains the lengthscale parameters for the input variables. For the input variables $H$, $S$ and $I$ correspond the parameters $\rho_1$, $\rho_2$ and $\rho_3$, respectively. The spatial coordinates input variables $S_x$ and $S_y$ are sharing the lengthscale parameter ($\rho_4=\rho_5$), such that the covariance function depends on the (Euclidean) distance between spatial coordinates.
Hyperparameter $\rho_d$ controls the smoothness of the covariance function or rate of decay of the correlation in the direction of the $d$th predictor, so that, the larger the $\rho_d$ the smoother the correlation function and the smoother the posterior functions for $\bm{b}_{k \bm{\cdot}}$; Although the scale of $\rho_d$ is dependent on the scale of the input data along the dimension $d$. The squared exponential covariance function prior is a stationary function with respect to the input variables and implies independence among the contributing effects of the different input variables.

\subsubsection*{Function value constraints (zero-order constraints)}
The modeling procedure has to ensure that the function values are zero at the subset $A=\{(t,i): t=1\}$ of starting time points, that is, the time-series must start in zero. The function values $f_{ti}$ in (\ref{eq_f}) evaluated at the subset $A$ results 
$$
\bm{f}_A=\bm{H}_{1 \bm{\cdot}} \bm{\beta} + \bm{W}_{1 \bm{\cdot}} \bm{b}= 0,
$$ 

\noindent where $\bm{H}_{1 \bm{\cdot}}=\{\bm{H}_{t \bm{\cdot}}:t = 1 \}$ and $\bm{W}_{1 \bm{\cdot}}= \{\bm{W}_{t \bm{\cdot}}:t = 1 \}$ correspond to matrices $H$ and $W$ evaluated at the subset $A$ of observations, respectively. 

This property can be specified by using virtual observations equal to zero at these points, $\bm{y}_{A}=\{y_{ti}:(t,i)\in A \}=0$, and the Dirac Delta function $\delta$ as an observational model for these observations,
\begin{eqnarray}\label{eq_regularnoisefree_splines}
\lefteqn{p(\bm{y}_{A}|\bm{f}_{A}) = } \nonumber \\
& \delta(\bm{y}_A - \bm{f}_{A}) = \delta(\bm{y}_A - (\bm{H}_{1 \bm{\cdot}} \bm{\beta} + \bm{W}_{1 \bm{\cdot}} \bm{b})).
\end{eqnarray}

\noindent While the rest of observations are considered to be contaminated with Gaussian noise $\sigma$,
\begin{eqnarray}\label{eq_regularnoise_splines}
\lefteqn{ p(\bm{y}_{-A}|\bm{f}_{-A},\sigma) } \nonumber \\ 
&& =\prod_{t,i\not\in A}\mathcal{N}(y_{ti}|f_{ti},\sigma^2),
\end{eqnarray}

\noindent where $\bm{y}_{-A}$ denote the dataset $\bm{y}$ without the subset $A$ of observations, $\bm{y}_{-A}=\{y_{ti}:(t,i)\notin A \}$. And $\bm{f}_{-A}$ denotes the latent function values $\bm{f}$ without the subset $A$ of observations, $\bm{f}_{-A}=\{f_{ti}:(t,i)\notin A \}$. 
%


\subsubsection*{Splines functional models with derivatives}

Color degradation is expected to be non-decreasing as a function of time and stabilize in the long term, that is, the time-series must be monotonically non-decreasing and stabilize eventually. These properties can be expressed in terms of the first order partial derivative of the penalized splines model for each time-series.  
The partial derivative of the penalized splines function of time-series $i$, $\bm{f}_{\bm{\cdot} i}$ in (\ref{eq_fsplines}), with respect to the time input variable takes the form
\begin{eqnarray}\label{eq_fsplines_dervi}
\bm{f}_{\bm{\cdot} i}' = \frac{\partial \bm{f}_{\bm{\cdot} i}}{\partial t} = \beta_{2i} +  \frac{\partial W}{\partial t} \bm{b}_{\bm{\cdot} i},
\end{eqnarray}
  
\noindent where $\frac{\partial W}{\partial t}$ is the partial derivative with respect to time of matrix $W$ of penalized spline function values (\ref{eq_W}),
$$
\frac{\partial W}{\partial t} = W'= \frac{\partial Z}{\partial t} \Omega^{-1/2}.
$$

\noindent In the previous equation, the element $(t,k)$ of the partial derivative of the matrix $Z$ is the partial derivative of the element $Z_{tk}$ (\ref{eq_Z}),
$$
\frac{\partial Z_{tk}}{\partial t} = 2(t-\kappa_k).
$$

\subsubsection*{Derivative constraints (first-order constraints)}
The derivative of function $f_{ti}$ in (\ref{eq_f}) is:
\begin{equation}\label{eq_derv_f}
f'_{ti} = \frac{\partial f_{ti}}{\partial t} = \bm{\beta}_{2i} + \frac{\bm{W}_{t \bm{\cdot}}}{\partial t} \bm{b}_{\bm{\cdot} i}
\end{equation}

In order to impose a saturation constraint for long term stabilization, virtual observations of the values of partial derivatives with respect to the time input variable equal to zero are considered. Most of the rock art painting systems analyzed so far show stabilization at or before the 10 minutes of MFS monitoring measurements. Therefore, virtual observations of the partial derivatives equal to zero are considered at the subset $B=\{(t,i): t=T\}$ of ending points of the set of time-series, 
$$
\bm{y}'_B= \frac{\partial \bm{y}_B}{\partial t} = 0,
$$ 

\noindent where $\bm{y}'_B$ and $\bm{y}_B$ denotes the partial derivative and function values, respectively, at the subset $B$ of points. A Dirac Delta function $\delta$ is considered as an observational model for these observations,
\begin{eqnarray}\label{eq_deriv_delta_splines}
\lefteqn{ p(\bm{y}'_B|\bm{f}'_B)  = } \nonumber \\
& \delta(\bm{y}'_B - \bm{f}'_B)  = \delta(\bm{y}'_B - (\bm{\beta}_{2 \bm{\cdot}} + \frac{\bm{W}_{T \bm{\cdot}}}{\partial t} \bm{b})),
\end{eqnarray}

\noindent where $\bm{f}'_B= \frac{\partial \bm{f}_B}{\partial t} = \bm{\beta}_{2 \bm{\cdot}} + \frac{\bm{W}_{T \bm{\cdot}}}{\partial t} \bm{b}=\{\beta_{2i} +  \frac{\partial W}{\partial t} \bm{b}_{\bm{\cdot} i}:t=T \}$ is the partial derivative of the functional models in (\ref{eq_fsplines_dervi}) at the subset points $B$.

On the other hand, the time-series are guaranteed to be non-decreasing as a function of time when the partial derivative of the spline function is positive. So, virtual observations of the sign of the partial derivatives equal to one at the subset $C$ of desired time points where to induce monotonicity are considered, 
$$
\bm{z}_C'= {\mathrm sign}\big( \frac{\partial \bm{f}_C}{\partial t}\big) = \{{\mathrm sign}(f'_{ti}):t,i \in C \} = 1.
$$

\noindent The probit function $\Phi: {\rm I\!R} \to (0,1)$ can be used as a likelihood for the signs of the partial derivatives \citep{Riihimaki_2011},
\begin{eqnarray}\label{eq_deriv_probit_splines}
%
p(\bm{z}_{C}|\frac{\partial \bm{f}_C}{\partial t}) =  \prod_{t,i\in C}\Phi\left(z_{ti} \cdot f'_{ti} \cdot \frac{1}{v} \right).
\end{eqnarray}

\noindent The function $\Phi$ is the standard Normal cumulative distribution function  and $v > 0$ is a parameter controlling the strictness of the constraint. When $v$ approaches zero ($v \to 0$), function $\Phi$ approaches the step function. We set the $v$ parameter equal to $10^{-4}$.

\subsubsection*{Likelihood function}

The joint likelihood of both regular observations $\bm{y}$ (noise observations (\ref{eq_regularnoise_splines}) and noise-free observations (\ref{eq_regularnoisefree_splines})) and derivative observations $\bm{y}'$ for monotonicity (\ref{eq_deriv_probit_splines}) and for long-term stabilization (\ref{eq_deriv_delta_splines}), given the parameters $\bm{\beta}$, $\bm{b}$ and $\sigma$, the function values $W$ and $H$, and the derivatives function values $\frac{W}{\partial t}$, results:

\begin{eqnarray} \label{eq_likeli}
\lefteqn{ p(\bm{y},\bm{y}'|\bm{\beta},\bm{b},H,W,\frac{W}{\partial t},\sigma) =} \nonumber \\
&& \prod_{t,i\not\in A}\mathcal{N}(y_{ti}|f_{ti},\sigma^2) \nonumber \\
&& \times \delta(\bm{y}_A - (\bm{H}_{1 \bm{\cdot}} \bm{\beta} + \bm{W}_{1 \bm{\cdot}} \bm{b}) \nonumber \\
&& \times \prod_{t,i\in C}\Phi\left(z_{ti} \cdot f_{ti}' \cdot \frac{1}{v} \right) \nonumber \\
&& \times \delta(\bm{y}'_B - (\bm{\beta}_{2 \bm{\cdot}} + \frac{\bm{W}_{T \bm{\cdot}}}{\partial t} \bm{b}))
\end{eqnarray}

\noindent with $f_{ti}$ and $f_{ti}'$ as in equations (\ref{eq_f}) and (\ref{eq_derv_f}), respectively.

\subsubsection*{Bayesian inference}
Bayesian inference is based on the posterior joint distribution of parameters given the data, which is proportional to the product of the likelihood and priors:
\begin{eqnarray}
p(\bm{\beta},\bm{b},\sigma^2,\theta|\bm{y},\bm{y}') \propto p(\bm{y},\bm{y}'|\bm{\beta},\bm{b},\sigma^2) p(\bm{\beta}) p(\bm{b}|\theta)p(\theta)p(\sigma^2) \nonumber
\end{eqnarray}

\noindent where $p(\bm{y},\bm{y}'|\bm{\beta},\bm{b},\sigma)$ is the likelihood of the model (\ref{eq_likeli}) and $p(\bm{b}|\theta)$ is the Gaussian process prior for the parameters $\bm{b}$ (\ref{eq_gpprior}). The vector $\theta$ contains the hyperparameters $\alpha$ and $\bm{\rho}$. We set normal prior distributions for the linear parameters $\bm{\beta}$, $p(\bm{\beta})=\mathcal{N}(\alpha|0,1)$, and positive half-normal prior distributions for the hyperparameters $\alpha$, $p(\alpha)=\mathcal{N}^+(\alpha|0,1)$, and $\sigma$, $p(\sigma)=\mathcal{N}^+(\sigma|0,1)$, and gamma distributions for the hyperparameters $\bm{\rho}$, $p(\rho_d)={\mathrm Gamma}(\rho_d|1,0.1)$ for all $d$. These correspond to weakly informative prior distributions based on prior knowledge about the magnitude of the parameters. 

Predictive inference of new output values $\bm{y}^\ast$ and ${\bm{y}'}^\ast$ for a new sequence of input values $X^\ast$ can be computed by sampling from the posterior joint predictive distribution $p(\bm{y}^\ast,{\bm{y}'}^\ast|\bm{y},\bm{y}')$.  

To posterior distribution of interest $p(\bm{\beta},\bm{b},\sigma^2,\theta|\bm{y},\bm{y}')$ is in general intractable. Hence, to estimate both the parameter posterior distribution and the posterior predictive distribution for this model, simulation and/or distributional approximations methods must be used. Simulation methods based on Markov chain Monte Carlo \citep{brooks_2011} are general sampling methods to obtain samples from the joint posterior distribution. 

The proposed model considers the same spatial covariance structure between splines coefficients belonging to different order of the splines knots. This requires $O(N^3)$ computation expense in covariance matrix inversion, where $N$ denotes the number of spatial locations. 

For large data sets where iterative simulation algorithms are too slow, modal and distributional approximation methods can be as efficient and approximate alternatives \citep{Gelman_2013}.

\subsection{Model checking, predictive performance and model selection}
Common procedures of checking normality and tendencies on the fitted residuals can be used. However, the \textit{posterior predictive checks}, which are also known as the \textit{leave-one-out probability integral transformation} (LOO-PIT), can be used as a more rigorous procedure in order to guarantee good model performance and ensure that the model is compatible with the observed data. They are based on computing the probability of new predictions to be lower or equal to their corresponding actual observations following a leave-one-out cross-validation procedure \citep{gelfand_1992,Gelman_2013}:
\begin{equation*} \label{eq:LOO-PIT}
\text{LOO-PIT}_{(t,i)\in D} = P(y^\ast_{(t,i)\in D} \leq y_{(t,i)\in D}),
\end{equation*}

\noindent where $D$ is the subset of observation indices of new predictions in the cross-validation. $y^\ast _{(t,i)\in D}$ are the new observations from the predictive distribution at the subset $D$, and $y_{(t,i)\in D}$ are the actual observations at this subset $D$. The similarity or provenance of these probabilities from standard uniform distributions endorses these probabilities with the desirable property of having the same interpretation across models, which implies good fit to the data and good prediction \citep{Bayarri_2000}. Using simulation methods for estimating and predicting a Bayesian model, computing the probability of a predicted value being minor the observed one is straightforward through the collection of simulated values. 

The \textit{expected log predictive density} (ELPD) evaluates, by averaging over all the steps in the leave-one-out cross-validation procedure, how far new data is from the model while taking the posterior uncertainties into account. It is based on the log-density of new data given the model \citep{vehtari_2012}:
\begin{equation*} \label{eq:lppd}
\text{ELPD} = \frac{1}{|D|} \sum_{(t,i)\in D} ln (p(y_{ti}|\boldsymbol{y}_{-D})),
\end{equation*}

\noindent where $|D|$ denotes the cardinality of the subset $D$ of observation indices in the cross-validation. $\boldsymbol{y}_{-D}$ denotes the dataset without the subset $D$ of observations, $\boldsymbol{y}_{-D}=\{y_{ti}:(t,i)\notin D \}$.

The \textit{mean square predictive error} (MSE) also evaluates, by averaging over all the steps in the leave-one-out cross-validation procedure, how far new data is from the model by using the distance (error) between the actual observation ($y_{ti}$) and the predictive mean ($\tilde{y}_{ti}$):
\begin{equation*} \label{eq:spe}
\text{MSE} = \frac{1}{|D|} \sum_{(t,i)\in D} (y_{ti} - \tilde{y}_{ti})^2.
\end{equation*}

Following a leave one observation out cross-validation scheme (CV1), the subset $D$ of observation indices will be just a single observation $(t,i)$, $D=\{(t,i)\}$. The LOO-PIT, ELPD, and MSE will be computed following CV1. The LOO-PIT will essentially be useful for model checking, ensuring that the model is compatible with the data. The ELPD and MSE will evaluate the predictive performance of individual observations $(t,i)$.

The end goal of this work is to predict complete color-fading time-series at new unobserved locations. In order to do that, a leave one location out cross-validation scheme (CV2) can be performed, where the subset of observation indices $D$ will be a complete time-series of a specific spatial location $i$, $D=\{(t,i): t\in \{1,\dots,T\}\}$. The statistics ELPD and MSE will be computed following CV2. Plots of predicted new time-series superimposed to their corresponding actual observations will be shown in order to visually evaluate the predictive performance. Model selection can be done by comparing the predictive performance between models using the ELPD and MSE statistics. The best model is who maximizes the ELPD and/or minimizes the MSE.

\section{Experimental results}\label{sec:results}

The posterior distributions and predictive distributions have been estimated by Hamiltonian Monte Carlo sampling methods \citep{neal_2011} using the Stan software \citep{Carpenter_2017}. Three simulation chains with different initial values have been launched. The convergence of the simulation chains was evaluated by the split-Rhat convergence diagnosis \citep{Gelman_2013} and the effective sample size of the chains. A value of 1 in the split-Rhat convergence statistic indicates convergence of the simulation chain, although conventionally accepted values of convergence would be between 1 and 1.1. In our case, a split-Rhat value lower than 1.05 has been obtained for all parameter simulation chains.

\begin{figure*}
  \centering
  \includegraphics[width=1\textwidth]{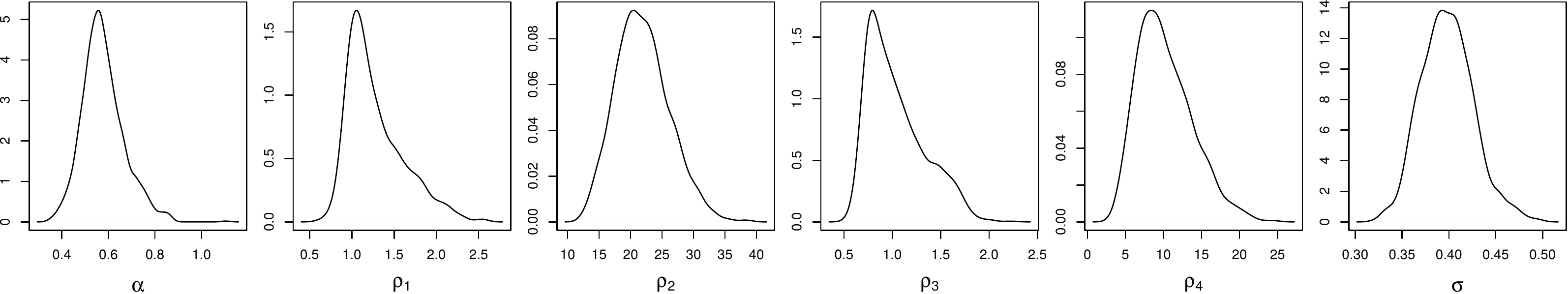}
  \vspace{-0.2cm}
\caption{Marginal posterior density distribution for the hyperparameters $\theta=(\alpha,\rho_d,\sigma), \hspace{0.1cm} d=1,\cdots,4$.}
\label{param}
\end{figure*}

The estimated marginal posterior distributions for parameters $\theta=(\alpha, \boldsymbol{\rho}, \sigma)$ are shown in Figure \ref{param}. $\alpha$ is the overall scale of the Gaussian process prior for the vectors of splines coefficients $\boldsymbol{b}_k$. $\sigma$ is the noise of the observations, and $\boldsymbol{\rho}=(\rho_1,\rho_2,\rho_3,\rho_4,\rho_5)$ are the lengthscale parameters associated with the input variables of the squared exponential covariance function in the Gaussian process priors. As mentioned above, the spatial coordinates input variables $S_x$ and $S_y$ are sharing the lengthscale parameter ($\rho_4=\rho_5$), such that the covariance function depends on the (Euclidean) distance between spatial coordinates. The lengthscale parameters $\rho_1$, $\rho_2$ and $\rho_3$ are associated to the input variables $H$, $S$ and $I$, respectively.

\begin{figure*}
  \centering
  \includegraphics[width=1\textwidth]{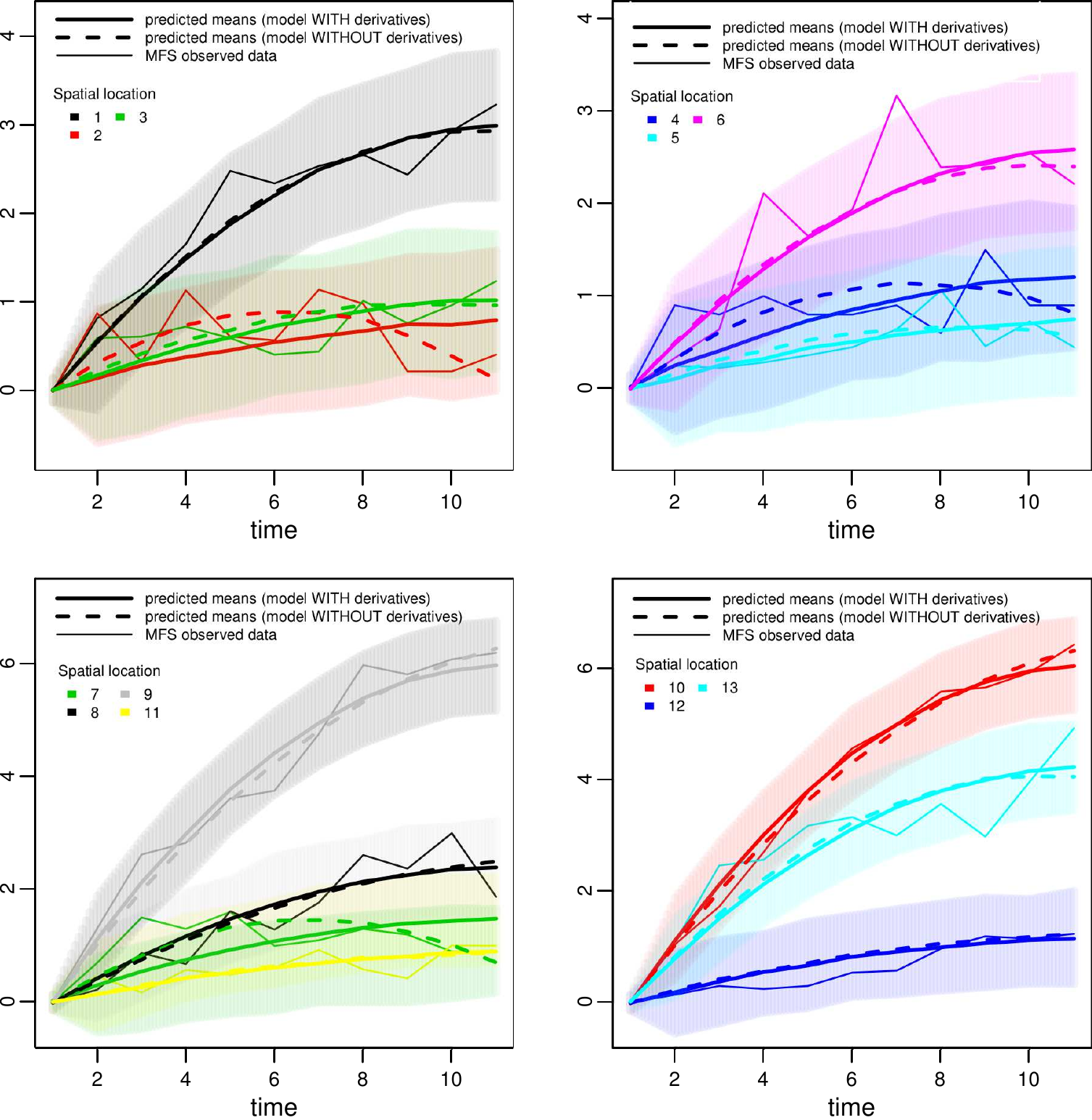} 
  \vspace{-0.5cm}
  \caption{Predictive distributions (means and 95\% pointwise credible intervals) of MFS time-series at spatial locations $i$, $p(\boldsymbol{y}^\ast_{\bm{\cdot} i}|\boldsymbol{y},\boldsymbol{y}')$, superimposed to the actual observations $\boldsymbol{y}_{\bm{\cdot} i}$. Predictive means for both models, with and without information about the derivatives, have been plotted. }
  \label{fitted}
\end{figure*}

\begin{figure*}
  \centering
  \includegraphics[width=1\textwidth]{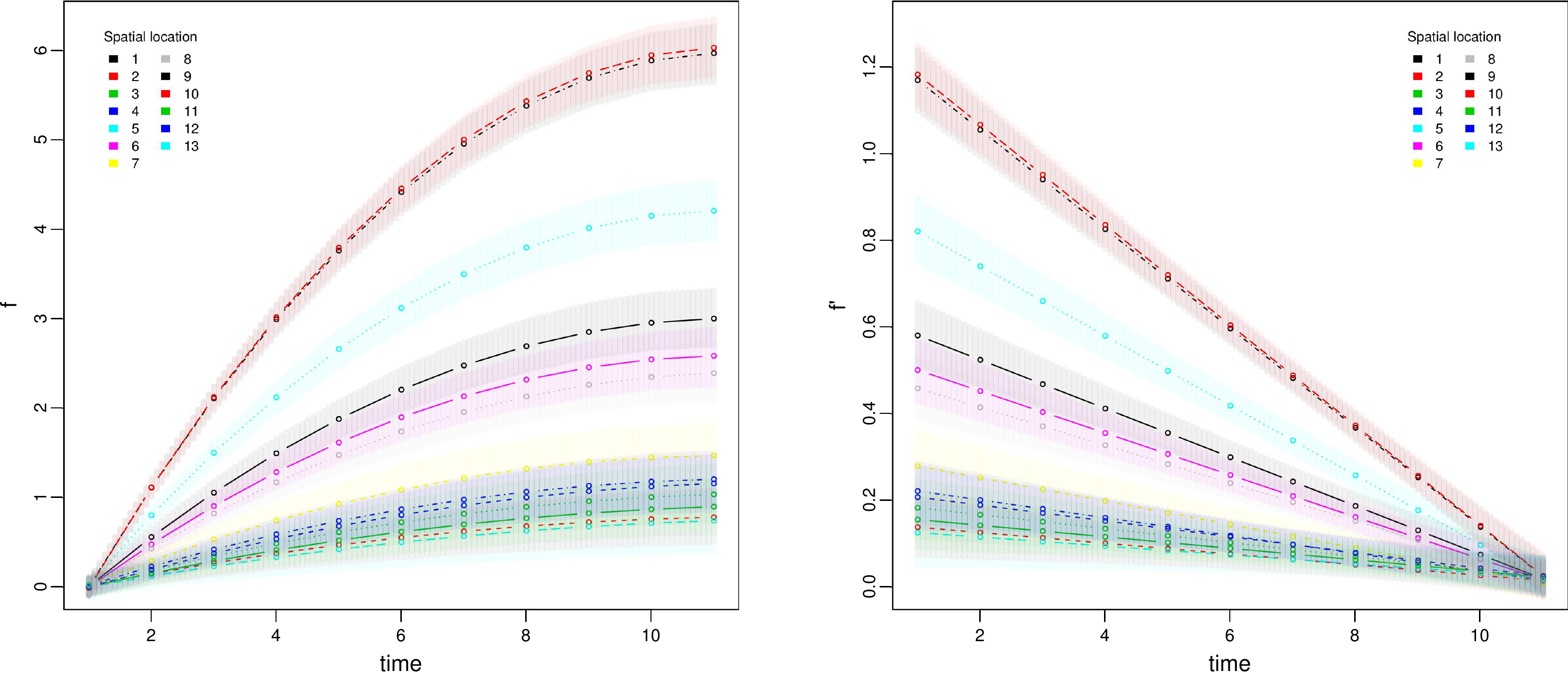} 
  \vspace{-0.5cm}
  \caption{Posterior distributions (means and 95\% pointwise credible intervals) of MFS time-series at spatial locations $i$, $p(\boldsymbol{f}_{\bm{\cdot} i}|\boldsymbol{y},\boldsymbol{y}')$ (left). Posterior derivatives means $\boldsymbol{f}'_{\bm{\cdot} i}$ and 95\% pointwise credible intervals of the derivative process $p(\boldsymbol{f}'_{\bm{\cdot} i}|\boldsymbol{y},\boldsymbol{y}')$. (right).}
  \label{fitted_derivatives}
\end{figure*}

Figure \ref{fitted} shows the means and 95\% pointwise credible intervals of predictive distributions, $p(\boldsymbol{y}^\ast|\boldsymbol{y},\boldsymbol{y}')$, evaluated over the observed data and plotted versus the time points and superimposed to the observed MFS data $\boldsymbol{y}$. Additionally, the means of the predictive distribution of the model without the inclusion of the derivatives, $p(\boldsymbol{y}^\ast|\boldsymbol{y})$, are also plotted for comparison. In Figure \ref{fitted_derivatives}, the means and 95\% pointwise credible intervals of the posterior distributions of the functions $\bm{f}$ and $\bm{f}'$, $p(\boldsymbol{f}|\boldsymbol{y},\boldsymbol{y}')$ and $p(\boldsymbol{f}'|\boldsymbol{y},\boldsymbol{y}')$, are plotted. 

\begin{figure}[h]
\centering
\subfigure{\includegraphics[scale=0.24]{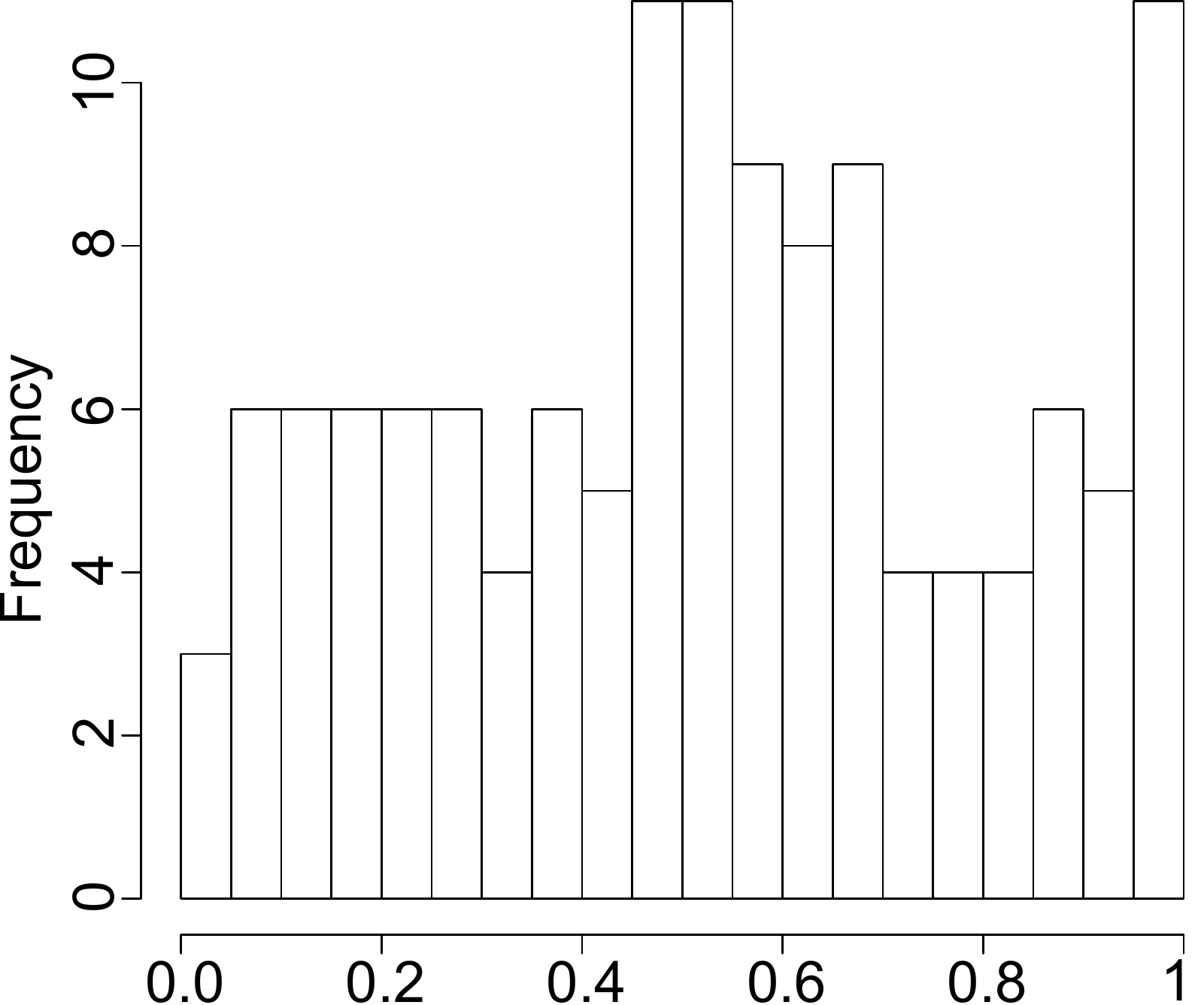}}
\hspace{0cm}
\subfigure{\includegraphics[scale=0.24]{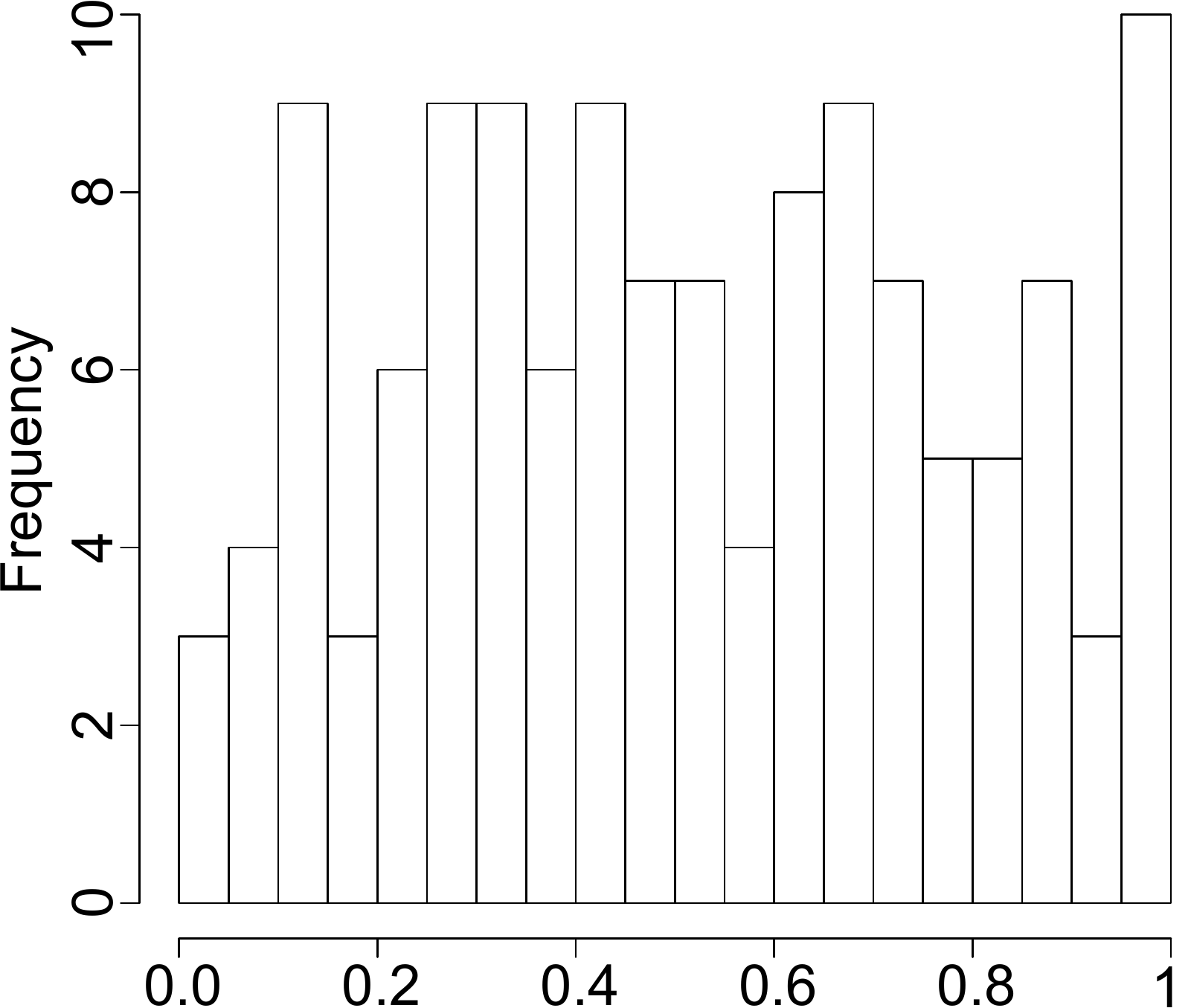}}
\caption{Histograms of the LOO probability integral transformation (LOO-PIT) for both the model with derivatives (left) and the model without derivatives (right).}
  \label{p_values}
\end{figure}

Figure \ref{p_values} shows the frequency histograms of the posterior predictive checks (LOO-PIT) by following the cross-validation scheme CV1, for both models, with and without derivatives.

\begin{figure*}
  \centering
  \includegraphics[width=1\textwidth]{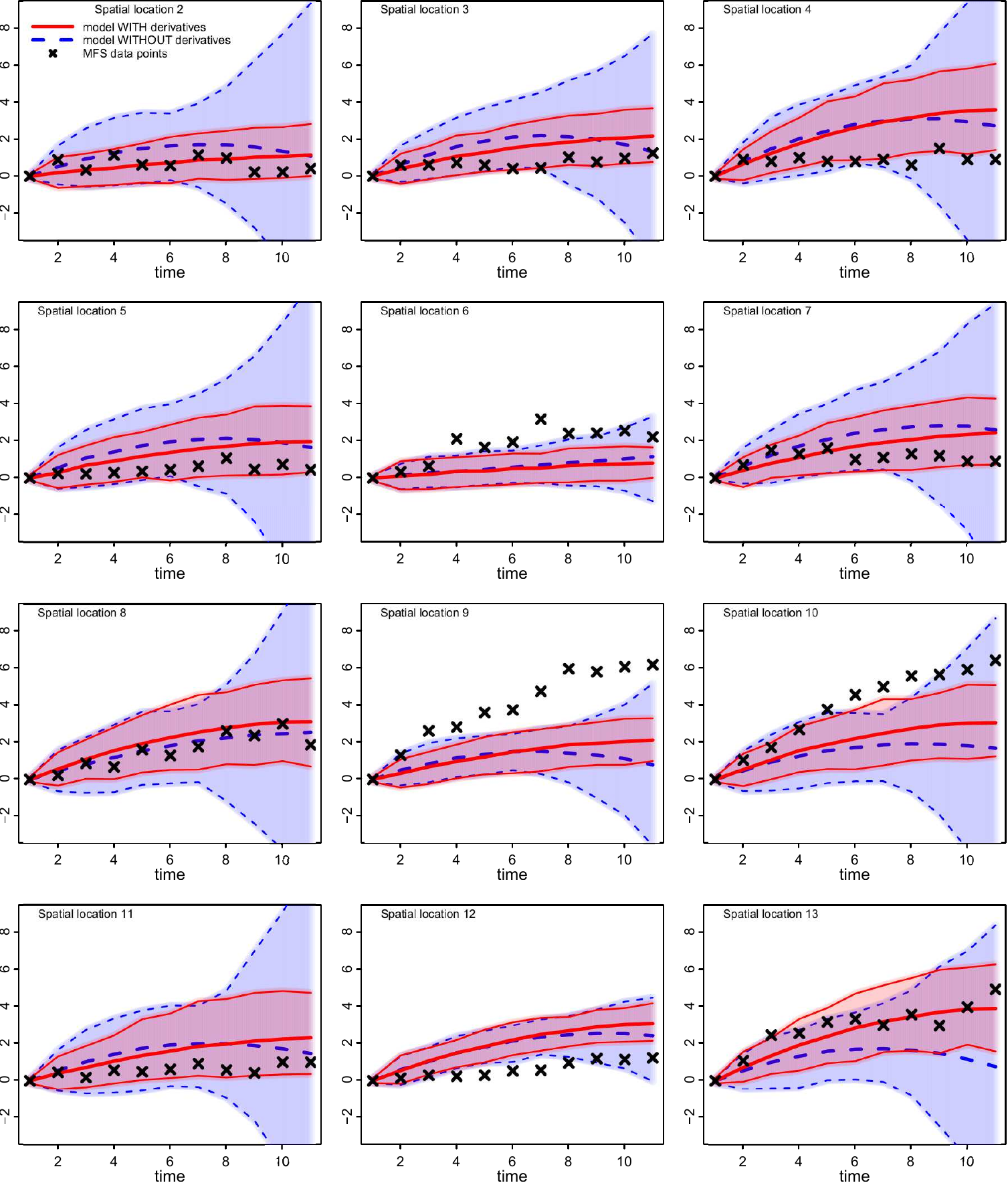} 
  \vspace{-0.5cm}
  \caption{Predicted distributions of time-series in the cross-validation scheme CV2, for both the model with derivatives and the model without derivatives. }
  \label{cross_validation}
\end{figure*}

Figure \ref{cross_validation} shows predictive distributions of the proposed model with derivatives, $p(\boldsymbol{y}^\ast_{\bm{\cdot} i}|\boldsymbol{y}_{\bm{\cdot} -i},\boldsymbol{y}'_{\bm{\cdot} -i})$, and of the model without derivatives, $p(\boldsymbol{y}^\ast_{\bm{\cdot} i}|\boldsymbol{y}_{\bm{\cdot} -i})$, for new time-series by following the cross-validation scheme CV2. In this way, predictive performance of complete new time series for unobserved locations is evaluated.  The actual data of the time-series $\boldsymbol{y}$ are also plotted to visually evaluate the predictions.

Table \ref{table} shows the ELPD and MSE statistics computed by following the two different cross-validation schemes, CV1 and CV2, and for both models, i.e. with and without derivatives. 

\begin{table}
\caption{\label{table}ELPD and MSE for both models, with and without derivative information, computed by the two cross-validation schemes.}

\centering
\fbox{
\begin{scriptsize}
\begin{tabular}{l|l|l|l}
 & & with derivative &  without derivative\\
 & & information & information \\
\hline
\multirow{2}{*}{CV1} & ELPD & -0.61 & -0.78 \\ 
& MSE & 0.13 & 0.14  \\
\hline
\multirow{2}{*}{CV2} & ELPD &  -11.70 & -33.39 \\ 
& MSE & 3.09  &  4.42  \\
\end{tabular}
\end{scriptsize}
}
\end{table}

\begin{figure*}[h]
\begin{minipage}{16cm}
\centering
  \includegraphics[width=1\textwidth]{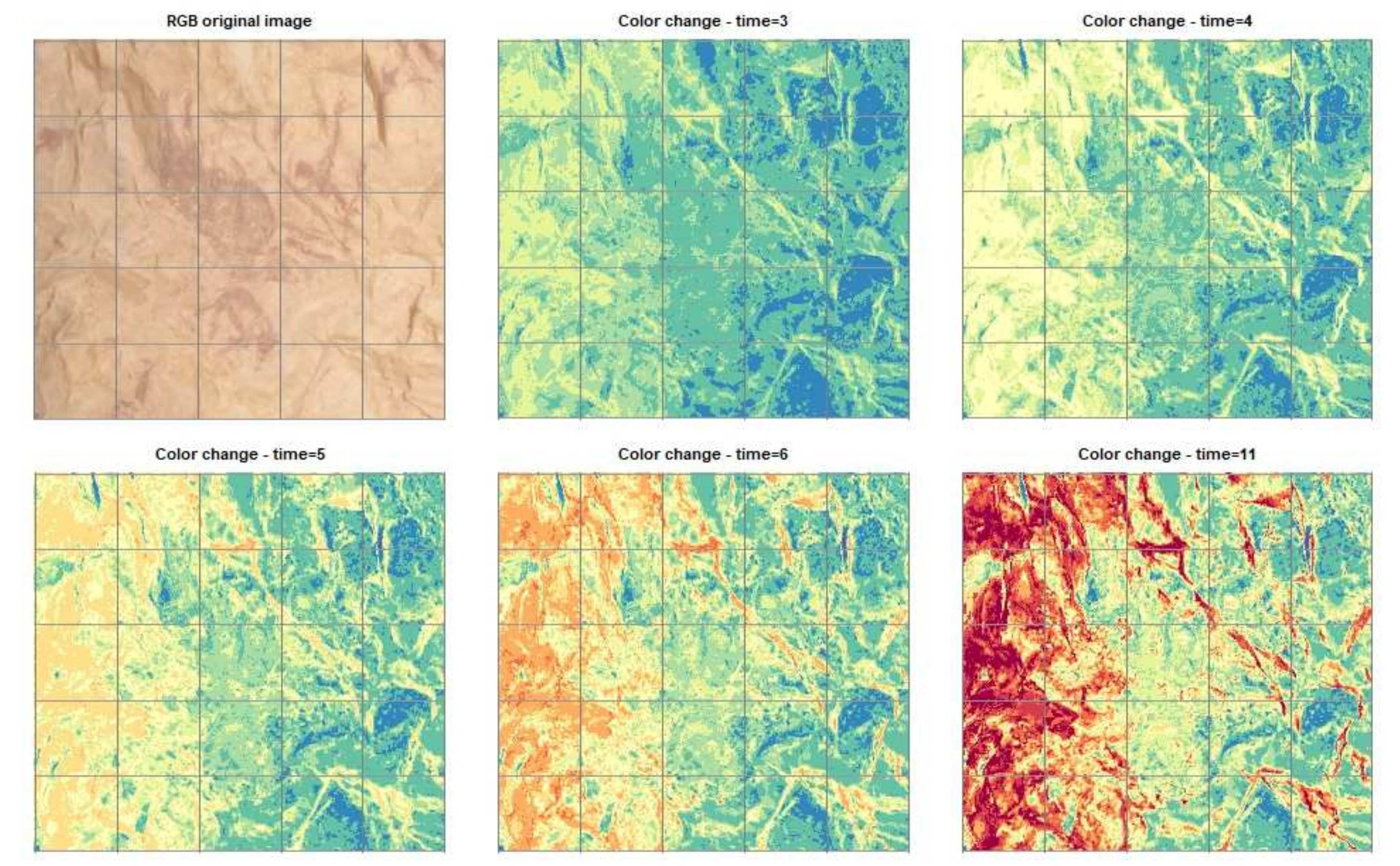}
  \vspace{-0.2cm}
  \caption{Spatial distribution over a continuous area of the rock art paintings image and the temporal evolution at time points $t=3$, $t=4$, $t=5$, $t=6$ and $t=11$, of mean color fading estimates.}
  \label{fading_images}
\end{minipage}
\begin{minipage}{0.4cm}
\centering
  \includegraphics[scale=0.5, trim= 187mm 0mm 0mm 0mm, clip]{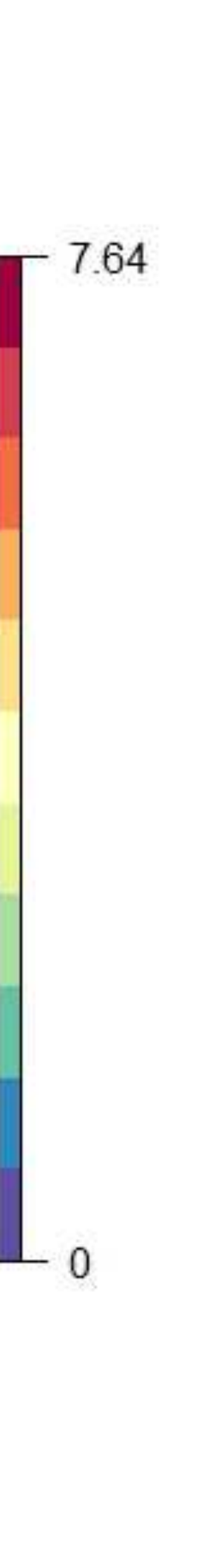}
  \vspace{-0.2cm}
\end{minipage}
\end{figure*}

In order to make spatial continuous maps of color fading estimates, predictions of color fading time-series, $p(\boldsymbol{y}^\ast _{\bm{\cdot} j}|\boldsymbol{y},\boldsymbol{y}')$, have been computed for all the spatial pixel locations $j$ of the rock art painting image (Figure \ref{observed_curves}). In Figure \ref{fading_images}, six images representing the spatial distribution and their evolution over time of those color fading estimates are shown. The images correspond to the time points $t=3$, $t=4$, $t=5$, $t=6$ and $t=11$, respectively. Color changes higher that aproximately 3.5 (deltaE*) are considered to be perceptible changes for the human eye \citep{Malacara_2011}.

\section{Discussion} \label{discussion}
The input variables $H$, $S$, $I$ and the spatial coordinates ($S_x$, $S_y$) have been previously standardized. The fact that the lengthscale parameters $\rho_1$ and $\rho_3$, corresponding to the variables $H$ and $I$, respectively, are relatively small (the mode of $\rho_1$ is 1.1 and the mode of $\rho_3$ is 0.8) indicates that the function is non-linear with those variables or that the rate of decay of the correlation is high so that variations in the input variables imply a quick decrease in the correlations allowing for the non-linear effects. The variables $S$ and spatial coordinates have larger lengthscales (modes of 20.0 and 8.9, respectively) so that the function depends on $S$ and spatial distances in a less non-linear way. Especially, the variable $S$ with a lengthscale of 20.0 contributes with a constant effect of one to the correlation, and implies a constant function to the Gaussian process, so being an irrelevant variable to the functions.  

Modeling and smoothing over the temporal dimension of the data has been achieved using penalized splines of order 3, as can be seen in Figure \ref{fitted}. 

The use of a noise-free pseudo observation model at the starting time points, $\{(t,i):t=1\}$, forces the predictive distributions $p(\boldsymbol{y}^\ast|\boldsymbol{y},\boldsymbol{y}')$ to be zero at those starting points (Figures \ref{fitted} and \ref{cross_validation}).

The inclusion of observations of the sign of the partial derivatives for all the data points, $\{{\mathrm sign}(y'_{ti})\geq0 \}$, has induced monotonicity (non-decreasing) through the whole time-series. In the same way, the inclusion of observations of the value of the partial derivative equal to zero at the ending time points, $\{ y'_{ti}=0:t=11 \}$, and a noise-free pseudo-observation model for these observations, has induced to reach a stationary state at the ending of the time-series. These constraints have to be considered either in the observed and predicting data points. 

In the Figure \ref{fitted_derivatives}-right, the derivative function values $p(\boldsymbol{f}'|\boldsymbol{y},\boldsymbol{y}')$ of the process for the observed data are plotted. Derivatives are always positive, approaching zero at the ending time points, which means the time-series are always non-decreasing and stabilizing in the long term. Derivative functions are linear, as can be seen in those plots, since functions have been modeled as quadratic splines-based functions.

Monotonicity and long term stabilization properties of the time-series were not ensured using the model without derivative information (Figures \ref{fitted_derivatives}-left and \ref{cross_validation}), hence, the proposed model with derivative information yields a better fit for the functions dynamics, improving their interpretability. In this sense, the analysis of the color fading time-series using a model without derivative information could not be properly done because the temporal degradation, specially at the last time points, is unrealistic.

The LOO probability integral transformation (LOO-PIT) for both models, with and without derivatives, have been computed following the leave-one-out cross-validation scheme CV1. The frequency histograms of these LOO-PIT shows similarities to uniform distributions for both models, which means good model performances and that the models are compatible with the observed data.

The cross-validation scheme CV2 based on leaving a  whole time-series out of the training data, has been carried out in order to evaluate the prediction performance of complete new time-series at new locations. Figure \ref{cross_validation} shows the predicted time-series at new locations following the cross-validation scheme CV2, using both models, with and without derivatives. Better predictive performance is appreciated for the model with derivatives. Predictions are closer to the actual data and predictive intervals are narrower, because monotonicity and saturation constraints improves and reduces the credible intervals of the predictions. The model without derivatives shows decreasing patterns on the functions which are not consistent with the prior knowledge. Credible intervals, especially at the last part of the splines-based functions (time-series), are getting much bigger. Here, it can be seen as imposing the monotonicity and saturation constraints on the splines functions improve considerably the credible intervals.

The ELPD and MSE statistics are mostly better for the model with derivatives, maximizing the ELPD and minimizing the MSE (Table \ref{table}), in both cross-validation schemes. Therefore, the results of these statistics confirm that the model with derivatives is closer to new data, either in terms of the expected log-density or the mean error.

Prediction sensitivity due to the short set of data available have been found. It can be seen in Figure 5 poor predictive performance when compared to the observed data. This is due to the high sensitivite of the model to leaving some data out.

The spatial distribution and evolution over time, of color fading estimates for an entire area is shown in Figure \ref{fading_images}. The spatial distribution over time of color fading values seems to be quite unvarying. This was expected since the time-series adjusted in the different locations have similar patterns, smooth, monotonic increasing and tending to stabilize in the long term (Figure \ref{fitted_derivatives}). Color fading values, specially when they are low values, like in this case (the maximun is of 7.64), are not worth being converted to the RGB color space and plotted as an image, because the color changes will be not visible in a RGB image. The best way is  plotting color deltaE* values (Figure \ref{fading_images}). The science of colorimetry argues that deltaE* values higher that aproximately 3.5 would be perceptible for the human eye looking at the real object.

The actual equivalency of the time points ($t=1,\dots,11$) used in MFS measurements in years depends on the hours and intensity of sunlight that affects the paintings on a changing everyday basis. Without proper monitoring of light, this equivalency is difficult to obtain. Although this aspect of the research was not considered in the current study, future work will include an evaluation of the location and geographical orientation of the paintings together with long-term monitoring of light and UV radiation with the aim of estimating the dose acting upon the paintings in years.

\section{Conclusion} \label{conclusion}
Color is an important aspect of documentation and conservation of historic and cultural heritage, such as rock art paintings. The MFS data are largely time-consuming and difficult to materialize, specially in these rock art systems, so an interpolation procedure is needed in order to make predictions of MFS data on unobserved locations. Furthermore, these measurements in these systems are contaminated with large fluctuations, so the consideration of constraints in the modeling in order to overcome possible modeling issues that may arise due to these large fluctuations is highly encouraged. 

We have formulated a model for correlating the MFS time-series of observations, with the addition of constraints related to their derivatives in order to fit some desired properties and minimize the effects of largely fluctuations in the original observations. Penalized splines of order 3 has been used for modeling the time-series. The time-series have been correlated through the spline coefficients using Gaussian processes models with multivariate predictors. In the practical case, we have been able to include colorimetric variables and the spatial coordinates variables in the covariance function, and demonstrated the colorimetric values being useful to correlate the MFS data. The contribution of the Euclidean distance (through the spatial coordinates variables) on this covariance structure is found to be quite weak.

Considering noise-free pseudo observations at the starting time points has allowed the functions to start at zero. The use of first order constraints related to the derivative of the functions have allowed the functions to fulfill the properties of being non-decreasing and stabilizing in the long term, and demonstrated better either in terms of predictive performance or application-specific interpretability. 

The computation requirements to invert the covariance matrix in the proposed correlated functional models have been reduced substantially compared to using a Gaussian process models with a spatio-temporal covariance function. The proposed general correlated functional models framework, where a complete covariance structure among splines coefficients is considered, requires $O\big((NK)^3\big)$ computational demand in covariance matrix inversion. Whereas spatio-temporal Gaussian processes require $O\big((NT)^3\big)$. Notice that the number of knots $K$ in spline-based models is usually much lower than the number of time points $T$ ($K\ll T$). In case a null covariance is considered between splines coefficients belonging to different spline knots, the computational expenses of the proposed model becomes $O(N^3 K)$. And furthermore, if the same spatial structure is considered for the splines coefficients belonging to different spline knots, as we implemented in the present paper, the computational expenses of the proposed model becomes $O(N^3)$.

Reliable color fading estimates evolution maps can be elaborated by means of using the proposed model with derivative information in comparison to the model without derivative information. 

Finally, multivariate covariance metrics and zero and first order constraints might be very hard to implement outside of a Bayesian framework and Gaussian process models. Gaussian process is flexible enough which allowed us to properly model this complex covariance structure of the time-series dependent on different input covariates. The Bayesian framework has allowed us to jointly use normal distributed observations with probit distributed observations of the sign of the partial derivatives, allowing to fulfill the determinants on the behavior of the functions.

\section*{Acknowledgments}

The authors gratefully acknowledge the support from the Spanish Ministerio de Econom\'ia y Competitividad to the projects HAR2014-59873-R and MTM2016-77501-P, as well as the grant PPIC-2014-001 funded by Consejer\'ia de Educaci\'on, Cultura y Deportes (JCCM) and FEDER. The authors want to express their gratitude to the General Directorate of Culture and Heritage, Conselleria d'Educaci\'o, Investigaci\'o, Cultura i Esport, Generalitat Valenciana for letting us access and carry out research at the archaeological site.

\bibliographystyle{biom}
\bibliography{references}

\begin{thebibliography}{}

\bibitem[\protect\citeauthoryear{Aguilera-Morillo, Durb{\'a}n, and
  Aguilera}{Aguilera-Morillo et~al.}{2017}]{aguilera2017prediction}
Aguilera-Morillo, M.~C., Durb{\'a}n, M., and Aguilera, A.~M. (2017).
\newblock Prediction of functional data with spatial dependence: a penalized
  approach.
\newblock {\em Stochastic environmental research and risk assessment} {\bf 31,}
  7--22.

\bibitem[\protect\citeauthoryear{Baladandayuthapani, Mallick, Young~Hong,
  Lupton, Turner, and Carroll}{Baladandayuthapani
  et~al.}{2008}]{Baladandayuthapani_2008}
Baladandayuthapani, V., Mallick, B.~K., Young~Hong, M., Lupton, J.~R., Turner,
  N.~D., and Carroll, R.~J. (2008).
\newblock Bayesian hierarchical spatially correlated functional data analysis
  with application to colon carcinogenesis.
\newblock {\em Biometrics} {\bf 64,} 64--73.

\bibitem[\protect\citeauthoryear{Banerjee, Carlin, and Gelfand}{Banerjee
  et~al.}{2014}]{banerjee_2014}
Banerjee, S., Carlin, B.~P., and Gelfand, A.~E. (2014).
\newblock {\em Hierarchical modeling and analysis for spatial data}.
\newblock Crc Press.

\bibitem[\protect\citeauthoryear{Bayarri and Berger}{Bayarri and
  Berger}{2000}]{Bayarri_2000}
Bayarri, M.~J. and Berger, J.~O. (2000).
\newblock P values for composite null models.
\newblock {\em Journal of the American Statistical Association} {\bf 95,}
  1127--1142.

\bibitem[\protect\citeauthoryear{Brezger and Steiner}{Brezger and
  Steiner}{2008}]{Brezger_2008}
Brezger, A. and Steiner, W.~J. (2008).
\newblock Monotonic regression based on bayesian p--splines: An application to
  estimating price response functions from store-level scanner data.
\newblock {\em Journal of business \& economic statistics} {\bf 26,} 90--104.

\bibitem[\protect\citeauthoryear{Brooks, Gelman, Jones, and Meng}{Brooks
  et~al.}{2011}]{brooks_2011}
Brooks, S., Gelman, A., Jones, G., and Meng, X.-L. (2011).
\newblock {\em Handbook of markov chain monte carlo}.
\newblock CRC press.

\bibitem[\protect\citeauthoryear{Carpenter, Gelman, Hoffman, Lee, Goodrich,
  Betancourt, Brubaker, Guo, Li, and Riddell}{Carpenter
  et~al.}{2017}]{Carpenter_2017}
Carpenter, B., Gelman, A., Hoffman, M.~D., Lee, D., Goodrich, B., Betancourt,
  M., Brubaker, M., Guo, J., Li, P., and Riddell, A. (2017).
\newblock Stan: A probabilistic programming language.
\newblock {\em Journal of Statistical Software} {\bf 76,} 1649--1660.

\bibitem[\protect\citeauthoryear{Cassar, Brimblecombe, and Nixon}{Cassar
  et~al.}{2001}]{Cassar_2001}
Cassar, M., Brimblecombe, P., and Nixon, T. (2001).
\newblock Technological requirements for solutions in the conservation and
  protection of historic monuments and archaeological remains.
\newblock {\em Working paper for the STOA Unit} .

\bibitem[\protect\citeauthoryear{Columbia, D., C., and Messier}{Columbia
  et~al.}{2013}]{Columbia_2013}
Columbia, M.~R., D., S.~G., C., H., and Messier, P. (2013).
\newblock The application of microfadeometric testing to mounted photographs at
  the indianapolis museum of art.
\newblock {\em Microscopy and Microanalysis} {\bf 19,} 1412--1413.

\bibitem[\protect\citeauthoryear{Crainiceanu, Ruppert, and Wand}{Crainiceanu
  et~al.}{2005}]{Crainiceanu_2005}
Crainiceanu, C.~M., Ruppert, D., and Wand, M.~P. (2005).
\newblock Bayesian analysis for penalized spline regression using winbugs.
\newblock {\em Journal of Statistical Software} {\bf 14,} 1--24.

\bibitem[\protect\citeauthoryear{Cressie and Huang}{Cressie and
  Huang}{1999}]{cressie1999classes}
Cressie, N. and Huang, H.-C. (1999).
\newblock Classes of nonseparable, spatio-temporal stationary covariance
  functions.
\newblock {\em Journal of the American Statistical Association} {\bf 94,}
  1330--1339.

\bibitem[\protect\citeauthoryear{Currie, Durban, and Eilers}{Currie
  et~al.}{2006}]{currie2006generalized}
Currie, I.~D., Durban, M., and Eilers, P.~H. (2006).
\newblock Generalized linear array models with applications to multidimensional
  smoothing.
\newblock {\em Journal of the Royal Statistical Society: Series B (Statistical
  Methodology)} {\bf 68,} 259--280.

\bibitem[\protect\citeauthoryear{De~Iaco, Myers, and Posa}{De~Iaco
  et~al.}{2002}]{de2002nonseparable}
De~Iaco, S., Myers, D.~E., and Posa, D. (2002).
\newblock Nonseparable space-time covariance models: some parametric families.
\newblock {\em Mathematical Geology} {\bf 34,} 23--42.

\bibitem[\protect\citeauthoryear{del Hoyo-Mel\'endez, Lerma, L\'opez-Montalvo,
  and Villaverde}{del Hoyo-Mel\'endez et~al.}{2015}]{del_hoyo_2015}
del Hoyo-Mel\'endez, J.~M., Lerma, J.~L., L\'opez-Montalvo, E., and Villaverde,
  V. (2015).
\newblock Documenting the light sensitivity of spanish levantine rock art
  paintings.
\newblock {\em ISPRS Ann Photogramm Remote Sens Spat Inf Sci} .

\bibitem[\protect\citeauthoryear{del Hoyo-Mel\'endez and Mecklenburg}{del
  Hoyo-Mel\'endez and Mecklenburg}{2010}]{del_hoyo_2010}
del Hoyo-Mel\'endez, J.~M. and Mecklenburg, M.~F. (2010).
\newblock A survey on the light-fastness properties of organic-based alaska
  native artifacts.
\newblock {\em J Cult Herit.} {\bf 11,} 493--499.

\bibitem[\protect\citeauthoryear{D\'iez-Herrero, Guti\'errez-P\'erez, and
  Lario}{D\'iez-Herrero et~al.}{2009}]{Diez-Herrero_2009}
D\'iez-Herrero, A., Guti\'errez-P\'erez, I., and Lario, J. (2009).
\newblock Analysis of potential direct insolation as a degradation factor of
  cave paintings in villar del humo, cuenca, central spain.
\newblock {\em Geoarchaeology} {\bf 24,} 450--465.

\bibitem[\protect\citeauthoryear{Feller, Johnston-Feller, and Bailie}{Feller
  et~al.}{1986}]{feller1986determination}
Feller, R.~L., Johnston-Feller, R.~M., and Bailie, C. (1986).
\newblock Determination of the specific rate constant for the loss of a yellow
  intermediate during the fading of alizarin lake.
\newblock {\em Journal of the American Institute for Conservation} {\bf 25,}
  65--72.

\bibitem[\protect\citeauthoryear{Ford}{Ford}{2011}]{Ford_2011}
Ford, B. (2011).
\newblock Non-destructive microfade testing at the national museum of
  australia.
\newblock {\em AICCM Bull.} {\bf 32,} 54--64.

\bibitem[\protect\citeauthoryear{Ford and Druzik}{Ford and
  Druzik}{2013}]{Ford_2013}
Ford, B. and Druzik, J. (2013).
\newblock Microfading: the state of the art for natural history collections.
\newblock {\em Collect. forum} {\bf 27,} 54--71.

\bibitem[\protect\citeauthoryear{Gelfand, Dey, and Chang}{Gelfand
  et~al.}{1992}]{gelfand_1992}
Gelfand, A.~E., Dey, D.~K., and Chang, H. (1992).
\newblock Model determination using predictive distributions with
  implementation via sampling-based methods.
\newblock Technical report, STANFORD UNIV CA DEPT OF STATISTICS.

\bibitem[\protect\citeauthoryear{Gelman, Carlin, Stern, Dunson, Vehtari, and
  Rubin}{Gelman et~al.}{2013}]{Gelman_2013}
Gelman, A., Carlin, J.~B., Stern, H.~S., Dunson, D.~B., Vehtari, A., and Rubin,
  D.~B. (2013).
\newblock {\em Bayesian data analysis}.
\newblock FL: CRC press, 3 edition.

\bibitem[\protect\citeauthoryear{Giles}{Giles}{1965}]{giles_1965}
Giles, C.~H. (1965).
\newblock The fading of colouring matters.
\newblock {\em J. Appl. Chem.} {\bf 15,} 541--550.

\bibitem[\protect\citeauthoryear{Giles, Johari, and Shah}{Giles
  et~al.}{1968}]{giles_1968}
Giles, C.~H., Johari, D.~P., and Shah, C.~D. (1968).
\newblock Some observations on the kinetics of dye fading.
\newblock {\em Textile Res. J.} {\bf 38,} 1048.

\bibitem[\protect\citeauthoryear{Giraldo, Delicado, and Mateu}{Giraldo
  et~al.}{2010}]{Giraldo_2010}
Giraldo, R., Delicado, P., and Mateu, J. (2010).
\newblock Continuous time-varying kriging for spatial prediction of functional
  data: an environmental application.
\newblock {\em Journal of agricultural, biological, and environmental
  statistics} {\bf 15,} 66--82.

\bibitem[\protect\citeauthoryear{Johnston-Feller, Feller, Bailie, and
  Curran}{Johnston-Feller et~al.}{1984}]{johnston1984kinetics}
Johnston-Feller, R., Feller, R.~L., Bailie, C.~W., and Curran, M. (1984).
\newblock The kinetics of fading: opaque paint films pigmented with alizarin
  lake and titanium dioxide.
\newblock {\em Journal of the American Institute for Conservation} {\bf 23,}
  114--129.

\bibitem[\protect\citeauthoryear{Kneib and Fahrmeir}{Kneib and
  Fahrmeir}{2006}]{kneib2006structured}
Kneib, T. and Fahrmeir, L. (2006).
\newblock Structured additive regression for categorical space--time data: A
  mixed model approach.
\newblock {\em Biometrics} {\bf 62,} 109--118.

\bibitem[\protect\citeauthoryear{Lee and Durb{\'a}n}{Lee and
  Durb{\'a}n}{2011}]{lee2011p}
Lee, D.-J. and Durb{\'a}n, M. (2011).
\newblock P-spline anova-type interaction models for spatio-temporal smoothing.
\newblock {\em Statistical Modelling} {\bf 11,} 49--69.

\bibitem[\protect\citeauthoryear{Malacara}{Malacara}{2011}]{Malacara_2011}
Malacara, D. (2011).
\newblock {\em Color vision and colorimetry: theory and applications}.
\newblock Washington: SPIE.

\bibitem[\protect\citeauthoryear{Neal}{Neal}{1999}]{Neal_1999}
Neal, R. (1999).
\newblock Regression and classification using gaussian process priors (with
  discussion).
\newblock {\em In Bernardo, J.M., Berger, J.O., Dawid, A.P., and Smith, A.F.M.,
  editors, Bayesian Statistics. Oxford University Press} {\bf 6,} 475--501.

\bibitem[\protect\citeauthoryear{Neal et~al\mbox{.}}{Neal
  et~al.}{2011}]{neal_2011}
Neal, R.~M. et~al. (2011).
\newblock Mcmc using hamiltonian dynamics.
\newblock {\em Handbook of Markov Chain Monte Carlo} {\bf 2,}.

\bibitem[\protect\citeauthoryear{Ramsay}{Ramsay}{1998}]{ramsay_1998}
Ramsay, J.~O. (1998).
\newblock Estimating smooth monotone functions.
\newblock {\em Journal of the Royal Statistical Society: Series B (Statistical
  Methodology)} {\bf 60,} 365--375.

\bibitem[\protect\citeauthoryear{Rasmussen}{Rasmussen}{2003}]{rasmussen_2003}
Rasmussen, C.~E. (2003).
\newblock Gaussian processes to speed up hybrid monte carlo for expensive
  bayesian integrals.
\newblock {\em Bayesian Statistics} {\bf 7,} 651--659.
\newblock Oxford University Press.

\bibitem[\protect\citeauthoryear{Rasmussen and Williams}{Rasmussen and
  Williams}{2006}]{Rasmussen_2006}
Rasmussen, C.~E. and Williams, C. K.~I. (2006).
\newblock {\em Gaussian Processes for Machine Learning}.
\newblock The MIT Press.

\bibitem[\protect\citeauthoryear{Reich, Fuentes, and Dunson}{Reich
  et~al.}{2011}]{Reich_2011}
Reich, B.~J., Fuentes, M., and Dunson, D.~B. (2011).
\newblock Bayesian spatial quantile regression.
\newblock {\em Journal of the American Statistical Association} {\bf 106,}
  6--20.

\bibitem[\protect\citeauthoryear{Riihim\"aki and Vehtari}{Riihim\"aki and
  Vehtari}{2010}]{Riihimaki_2011}
Riihim\"aki, J. and Vehtari, A. (2010).
\newblock Gaussian processes with monotonicity information.
\newblock In JMLR, editor, {\em Proceedings of the 13th International
  Conference on Artificial Intelligence and Statistics (AISTATS)}, volume~9.

\bibitem[\protect\citeauthoryear{Shively, Sager, , and Walker}{Shively
  et~al.}{2009}]{shively_2009}
Shively, T.~S., Sager, T.~W., , and Walker, S.~G. (2009).
\newblock A bayesian approach to non-parametric monotone function estimation.
\newblock {\em Journal of the Royal Statistical Society (Series B)} {\bf 71,}
  159--175.

\bibitem[\protect\citeauthoryear{Vehtari, Ojanen, et~al\mbox{.}}{Vehtari
  et~al.}{2012}]{vehtari_2012}
Vehtari, A., Ojanen, J., et~al. (2012).
\newblock A survey of bayesian predictive methods for model assessment,
  selection and comparison.
\newblock {\em Statistics Surveys} pages 142--228.

\bibitem[\protect\citeauthoryear{Whitmore, Bailie, and Connors}{Whitmore
  et~al.}{2000}]{Whitmore_2000}
Whitmore, P., Bailie, C., and Connors, S.~A. (2000).
\newblock Micro-fading tests to predict the result of exhibition progress and
  prospects.
\newblock {\em Stud. Conserv} {\bf 45,} 200--205.

\bibitem[\protect\citeauthoryear{Whitmore, Pan, and Bailie}{Whitmore
  et~al.}{1999}]{Whitmore_1999}
Whitmore, P.~M., Pan, X., and Bailie, C. (1999).
\newblock Predicting the fading of objects: Identification of fugitive
  colorants through direct nondestructive lightfastness measurements.
\newblock {\em J Am Inst Conserv} {\bf 38,} 395--409.

\bibitem[\protect\citeauthoryear{Wood}{Wood}{2003}]{wood2003thin}
Wood, S.~N. (2003).
\newblock Thin plate regression splines.
\newblock {\em Journal of the Royal Statistical Society: Series B (Statistical
  Methodology)} {\bf 65,} 95--114.

\end{thebibliography}

\end{document}